\documentclass[aps,prb,preprint,showpacs,groupedaddress]{revtex4}

\usepackage{amsmath}
\usepackage{amssymb}
\usepackage{graphicx}
\usepackage{dcolumn}
\usepackage{bm}
\usepackage{subfigure}

\begin{document}

\title{Atomic kinetic energy, momentum distribution and structure of solid neon at zero-temperature}

\author{C. Cazorla$^{1,2}$ and J. Boronat$^{3}$}
\affiliation{
$^{1}$London Centre for Nanotechnology, UCL, London WC1H OAH, UK \\
$^{2}$Department of Physics and Astronomy, UCL, London WC1E 6BT, UK \\
$^{3}$Departament de F\'isica i Enginyeria Nuclear, UPC, Campus Nord B4-B5, Barcelona E-08034, Spain
}
\

\begin{abstract}

We report on the calculation of the ground-state atomic kinetic energy, $E_{k}$, and momentum distribution of solid Ne
by means of the diffusion Monte Carlo method and Aziz HFD-B pair potential. This approach is shown to  
perform notably for this crystal since we obtain very good agreement with respect to experimental thermodynamic data. 
Additionally, we study the structural properties of solid Ne at densities near the equilibrium by estimating 
the radial pair-distribution function, Lindemann's ratio and atomic density profile around the positions of the
perfect crystalline lattice.
Our value for $E_{k}$ at the equilibrium density is $41.51(6)$~K, which agrees perfectly with the recent prediction made
by Timms {\it et al.}, $41(2)$~K, based on their deep-inelastic neutron scattering experiments carried out over the temperature range $4 - 20$~K,
and also with previous path integral Monte Carlo results obtained with the Lennard-Jones and Aziz HFD-C2 atomic pairwise
interactions.  
The one-body density function of solid Ne is calculated accurately and found to fit perfectly, within statistical uncertainty,
to a Gaussian curve.
Furthermore, we analyze the degree of anharmonicity of solid Ne by calculating some of its microscopic ground-state properties within
traditional harmonic approaches.   
We provide insightful comparison to solid $^4$He in terms of the Debye model, in order to size the relevance of anharmonic effects in Ne.

\end{abstract}


\maketitle

\section{Introduction}

 Noble gases like He, Ne, Xe and Ar, have been intensively studied during the last decades,
 both experimentally and theoretically.~\cite{pollack64,klein76,hansen76}
 Due to their simple electronic closed-shell structure, they appear to be affordable many-body
 systems where to carry out feasible quantum computations
 and test novel methods of calculation. Even though most of them are regarded as classical systems, 
 microscopic quantum approaches are required to understand the behavior of the lighter ones, He and Ne,
 at low temperatures. As it is well-known, He is the most representative of the quantum many-body systems.
 Unique features like Bose-Einstein condensation and superfluidity take
 place in the liquid at few K and recently several experimental groups have
 detected superfluid signal in the solid phase in the mK range,~\cite{kim04,rittner06}
 a signal that in the homogeneous crystal has been ruled out by accurate theoretical calculations.~\cite{profkofev05,ceperley04}
 Moreover, the atomic momentum distribution, $n({\bf k})$, of $^4$He differs significantly from  
 those of classical systems leading to a non-Gaussian $n({\bf k})$ curve sharply peaked around 
 $k = 0$.~\cite{azuah95,azuah97,diallo04}
 The reasons for those phenomena to happen in helium are the light mass of the atoms, bosonic nature of the system
 and weakness of the interparticle interactions.
 On the other side, Ne has long attracted the interest of condensed-matter scientists since it is an 
 intermediate quantum system which provides valuable physical insight when compared to other quantum and classical
 systems. Indeed, the De Boer quantum parameter~\cite{deboer49} defined as
        \begin{equation}
        \label{eq:deboerne}
        \Lambda^{*} = \frac{h}{\sqrt{m\epsilon\sigma^{2}}}~, 
        \end{equation}
 where $m$, $\epsilon$ and $\sigma$ are the atomic mass, energy scale of the atomic interactions and typical interatomic distance of the system,
 respectively, amounts to $0.54$ in Ne ($2.50$ in $^{4}$He) while in Ar and other heavier noble gases,
 where classical behavior is expected, it drops significantly to zero. 
 Essentially, the quantum character of liquid and solid Ne is evidenced on their atomic kinetic energies and momentum
 distributions, which differ appreciably from the predictions made by Classical Statistical Mechanics.
 Accordingly, anharmonic effects in the crystal may develop important at low temperatures due to the large 
 zero-point motion of the atoms.~\cite{hakim88}

In this work, we study solid Ne at zero temperature by means of the diffusion Monte Carlo method (DMC)~\cite{hammond94,guardiola98,ceperley79} 
and the Aziz HFD-B pair potential.~\cite{aziz89}
Our approach is microscopic and \emph{exact} in the sense that the total and partial ground-state energies of the crystal 
may be calculated within statistical uncertainty only.  
There are burdens of theoretical and experimental papers dealing with the thermodynamics and lattice dynamics of solid neon,
however, numerical results for the atomic kinetic energy are not so abundant. 
By the beginning of the 60's, Bernades~\cite{bernades58} and Nosanow \emph{et al.}~\cite{nosanow62} were the first in attempting
to estimate $E_{k}$ theoretically. 
They used uncorrelated single-particle wave functions within the variational and Hartree approaches, respectively, and
arrived at reasonable values not too far from present-day calculations; however, the binding energies that they reported     
were in significant disagreement with experimental data. These results made evident the need of 
improved theoretical schemes where to account for the atomic correlations in Ne. 
Few years after Bernades and Nosanow works, Koehler estimated $E_{k} = 42.6$~K by means of the Self Consistent Phonon 
approach (SCP), improving mildly the agreement with experiments.~\cite{koehler66} 

On the experimental side, however, it has not been until the beginning of the 80's, with the development 
of the deep-inelastic neutron scattering technique (DINS), that direct measurement of $E_{k}$ in the condensed phases of matter has become
accessible. Peek \emph{et al.} performed the first measurements in solid Ne, covering the 
temperature interval $4.5 - 26.5$~K.~\cite{peek92}
The authors of the first study reported $E_{k} = 49.1\pm 2.8$~K for the ground-state kinetic energy and, because of the large discrepancies
with respect to calculations based on harmonic models, they suggested substantial anharmonic effects in solid Ne.
  
Reassuringly, few years after Peek \emph{et al.}'s measurements,~\cite{peek92} theoretical estimations by Asger and Usmani,~\cite{asger94}
who used a perturbational approach based on a Wigner-Kirkwood high-temperature expansion with the Lennard-Jones (L-J) and Aziz HFD-C2~\cite{aziz83} 
pair potentials, amounted to $E_{k} \sim 49$~K at temperatures near $10$~K.
Regardless, previous to Asger {\it et al.}'s results,~\cite{asger94} Cuccoli and co-workers~\cite{cuccoli93} arrived at kinetic energies $\sim 7$~K
below Peek's results, based on the full quantum approach path integral Monte Carlo (PIMC) and the L-J interaction.
The authors of this work suggested that their disagreement with Peek's results could be in part due to the
oversimplification of the atomic interactions made by the adopted potential.
Subsequently, Timms \emph{et al.}~\cite{timms96} performed a series of new low-temperature DINS
experiments in solid Ne at high momentum tranfers with an improved experimental set-up. 
They found very good agreement with Cuccoli {\it et al.}~\cite{cuccoli93} and also with Ceperley and Boninsegni,~\cite{timms96} who performed an exhaustive
PIMC study of the crystal at low temperatures using both L-J and HFD-C2 pair potentials.  
In addition, a recent theoretical study by Neumann and Zoppi, in which computational techniques and interatomic potentials
similar to those of Ref.~\onlinecite{timms96} are used, comes to reinforce the accuracy of Timms' data.~\cite{neumann02} 
Very recently, Timms \emph{et al.}~\cite{timms03} have reported new additional DINS measurements in solid Ne
performed within the temperature range $4 - 20$~K.
By doing this, they complement their previous results and provide a truster way to infer the value of $E_{k}$ in the ground state,
which by means of extrapolation of the excess kinetic energy turns out to be $41(2)$~K. 

In the present work, we report quantum Monte Carlo results of the equation of state and other thermodynamic 
properties of solid Ne over a range of densities near equilibrium ($ -1.2 \le P \le 6$~Kbar), and find overall
excellent agreement with experimental data. 
Structural properties of the crystal, namely the radial pair-distribution function, $g(r)$, atomic density profile
around the positions of the perfect crystalline lattice (sites) and Lindemann ratio, are also provided.
Remarkably, we estimate accurately the atomic kinetic energy of the crystal at its equilibrium density by means of the
pure estimator technique within DMC.~\cite{liu74,reynolds86,casulleras95} 
Our result, $E_{k} = 41.51(6)$~K, is in very good agreement with the recent prediction of 
Timms \emph{et al}.~\cite{timms03} We have also calculated the ground-state atomic momentum distribution 
$n({\bf k})$ of solid Ne and it is found to fit perfectly to a Gaussian within the statistical uncertainty.

Additionallly, we have analyzed the degree of anharmonicity of solid Ne in its ground state.
With this aim, we have computed the atomic kinetic energy and mean squared displacement within the Self Consistent Average
Phonon (SCAP) approach,~\cite{shukla81,paskin82} which is a simplified version of the Self Consistent Phonon method~\cite{glyde94}
that has proved successful in reproducing a deal of thermodynamic properties of rare gas solids. We find the SCAP
results are not in full agreement with the quantum DMC ones, thus revealing this approach might not allow for an accurate description
of Ne at the microscopic level. 
In a further step, we devise an harmonic model based on the HFD-B potential in which the interaction between  
particles depend on their relative distances, equilibrium positions and the force constant field 
(second derivatives of the potential energy evaluated in the perfect crystal configuration).
By using DMC, we calculate the total and kinetic energies associated to this model
and find significant discrepancies with respect to the full HFD-B results. 
According to these outcomes, solid Ne may be regarded as a \emph{moderate} anharmonic crystal since, contrarily to what is
observed in solid $^{4}$He, its $n({\bf k})$ does not deviate appreciably from the Gaussian pattern.

The remainder of this article is as follows. In Sec.~\ref{sec:techniques}, we describe the computational techniques and 
models that have been used on this study. Next, in Sec.~\ref{sec:results}, we present our results and compare to previous
experimental and theoretical data. In Sec.~\ref{sec:discussion}, we finalize by summarizing the main conclusions and 
giving some general remarks.

\section{Techniques and Model}
\label{sec:techniques}

\subsection{Diffusion Monte Carlo}
\label{subsec:dmc}

DMC is a zero-temperature method which provides the exact ground-state energy of the many-boson
interacting systems within some statistical errors.~\cite{hammond94,guardiola98,ceperley79} 
This technique is based on a short-time approximation for the Green's function corresponding 
to the imaginary time-dependent Schr${\rm \ddot{o}}$dinger equation, which is solved 
up to a certain order of accuracy within an infinitesimal interval $\Delta \tau$. 
Despite this method is algorithmically simpler than domain Green's function Monte Carlo,~\cite{ceperley79,kalos74} it presents some
 $\left(\Delta \tau\right)^{n}$
bias coming from the factorization of the imaginary time propagator $e^{-\frac{\Delta\tau}{\hbar}{\rm H}}$.
Nevertheless, our implementation of DMC is quadratic,~\cite{chin90} hence the control of the time-step bias is efficiently controlled since 
the required $\Delta\tau \to 0$ extrapolation is nearly eliminated by choosing a sufficiently small time step.
The Hamiltonian ${\rm H}$, describing our system is 
\begin{equation}
\label{eq:hamiltonian}
{\rm H} = - \frac{\hbar^{2}}{2m_{\rm Ne}} \sum_{i=1}^{N} \nabla^{2}_{i} +
       \sum_{i<j}^{N} V_{2}(r_{ij})~,
\end{equation}  
where $m_{\rm Ne}$ is the mass of a Ne atom, $r_{ij}$ the distance between atoms composing an $i$,$j$ pair
and $V_{2}(r_{ij})$ the interatomic interaction that we have chosen as the Aziz HFD-B potential.~\cite{aziz89}
The corresponding Schr${\rm \ddot{o}}$dinger equation in imaginary time ($it \equiv \tau$), 
\begin{equation}
\label{eq:schrodinger}
-\hbar\frac{\partial \Psi({\bf R},\tau)}{\partial \tau}= \left({\rm H}-E\right)\Psi({\bf R},\tau) 
\end{equation}
with $E$ an arbitrary constant, can be formally solved by expanding the
solution $\Psi({\bf R}, \tau)$ in the basis set of the energy eigenfunctions $\{\Phi_{n}\}$. It turns out that
$\Psi({\bf R}, \tau)$ tends to the ground-state wave function $\Phi_{0}$ of the system for an infinite imaginary time as well as the
expected value of the Hamiltonian tends to the ground-state value $E_{0}$. 
The hermiticity of the Hamiltonian guarantees the equality   
\begin{equation}
\label{eq:groundstate1}
E_{0} = \frac{\left<\Phi_{0}|{\rm H}|\Phi_{0}\right>}{\left<\Phi_{0}|\Phi_{0}\right>}=
\frac{\left<\Phi_{0}|{\rm H}|\psi_{T}\right>}{\left<\Phi_{0}|\psi_{T}\right>} = \langle {\rm H} \rangle_{DMC} ~, 
\end{equation}
where $\psi_{T}$ is a convenient trial wave function which depends on the atomic coordinates of the system
${\bf R}\equiv \{ {\bf r_{1}}, {\bf r_{2}},...,{\bf r_{N}} \}$.
Consequently, the ground-state energy of the system can be computed by calculating the integral
\begin{equation}
\label{eq:integral}
\langle {\rm H} \rangle_{DMC}= \lim_{\tau \to\infty} \int_{V} E_{L}\left({\bf R}\right) f\left({\bf R},\tau\right) d{\bf R} \quad ,
\end{equation} 
where $f\left({\bf R},\tau\right)=\Psi\left({\bf R},\tau\right)\psi_{T}\left({\bf R}\right)$,
and $E_{L}\left({\bf R}\right)$ is the local energy defined as
$E_{L}({\bf R})= {\rm H}\psi_{T}\left({\bf R}\right)/\psi_{T}\left({\bf R}\right)$. 
The introduction of $\psi_{T}\left({\bf R}\right)$ in $f\left({\bf R},\tau\right)$ is known as importance sampling and   
it certainly improves the way in which integral (\ref{eq:integral}) is computed 
(for instance, by imposing $\psi_{T}\left({\bf R}\right)=0$ when $r_{ij}$ is smaller than the core distance
of the interatomic interaction).

In this work, all the operators diagonal in real-space which do not commute with the Hamiltonian, that is  $[ {\rm H}, \hat{O}] \neq 0$,
have been sampled with the pure estimator technique.~\cite{liu74,reynolds86,casulleras95}
With this method, essentially, the possible bias induced by $\psi_{T}$ in the mixed estimator $\left<\Phi_{0}|\hat{O}|\psi_{T}\right>$
are removed by proper weighting of the configurations along the simulation.

\subsection{Trial wave function and pair potential}
\label{subsec:twfpp}

We have modeled solid Ne by assuming point-like atoms interacting via a radial pair-wise potential
and with equilibrium positions distributed according to the fcc structure. Neon is observed to remain stable 
in the fcc structure up to pressures of $1100$~Kbar and at ambient temperature,~\cite{hemley89} therefore, no other configuration 
apart from this has been considered in the present study.
The potential chosen for the interatomic interactions
is the semi-empirical Aziz HFD-B one,~\cite{aziz89} which has proved excellent in 
reproducing some of the macroscopic and microscopic properties of Ne over a wide range of temperature and
pressure,~\cite{drummond06} and appears to be more realistic
than the Aziz HFD-C2~\cite{aziz83} and Lennard-Jones (L-J) models at short distances. 
Explicitly,
\begin{equation}
V(r)= \epsilon  \Theta (x)~,
\end{equation}
where
\begin{displaymath}
\Theta (x) = A \exp\left(-\alpha x + \beta x^{2}\right) 
    -F(x) \left( \frac{C_{6}}{x^{6}} +
     \frac{C_{8}}{x^{8}} + \frac{C_{10}}{x^{10}}\right)~,
\end{displaymath}
and
\begin{displaymath}
F(x)= \left\{ \begin{array}{ll}
      \exp \left[ -\left( \frac{D}{x} - 1 \right)^{2}\right] &  x < D   \\
      1   &     x \geq D ~.
\end{array} \right.	                  
\end{displaymath}
The value of the parameters of the potential are $ A= 895717.95$~,  $\alpha = 13.86434671$, $D = 1.36 $, $r_{m} = 3.091$~\AA~, $\beta = -0.12993822$~,
$\epsilon = 42.25$~K~, $C_{6} = 1.21317545$~,  $C_{8} = 0.53222749$~ and $C_{10} = 0.24570703$~, with $x \equiv r / r_{m}$~.  
It is known that, upon high pressure the introduction of additional terms in the effective atomic potentials of rare gases 
are required to account for many-body effects taking place on them;
for instance, in solid Ar this limit is posed around $50$~Kbar.~\cite{grimsditch86,lotrich97} 
This circumstance, however, does not affect the reliability of the results that we are to present in short, 
since the pressure range involved in our simulations is
$-1.2 \le P \le 6$~Kbar.  

Regarding the trial wave function chosen for importance sampling, $\psi_{T}$, we have adopted the  
extensively used and tested Nosanow-Jastrow model,~\cite{nosanow64,hansen68,hansen69}  
\begin{eqnarray}
\label{eq:nosanojasne}
\psi_{T}\left({\bf r_{1}},{\bf r_{2}},...,{\bf r_{N}}\right) =
\prod_{i\neq j}^{N} {\rm f_{2}}(r_{ij}) \prod_{i=1}^{N}{\rm g_{1}}(|{\bf r_{i}}-
{\bf R_{i}}|) ~,
\end{eqnarray}
with ${\rm f_{2}}(r) = e^{-\frac{1}{2}\left(\frac{b}{r}\right)^{c} }$ and ${\rm g_{1}}(r) = e^{-\frac{1}{2}ar^{2}}$~.
The best parameter values are $a = 6.5$~\AA$^{-2}$, $b = 4.0$~\AA ~ and $c = 5.0$~, optimized using the 
variational Monte Carlo method. Their dependence with the pressure is small and therefore neglected for its use
on the DMC simulations. 
The first factor in $\psi_{T}$ accounts for the correlations between particles induced by the interactions, while
the second enforces the atomic ordering within the system by attaching each particle to one site of the perfect
lattice through a Gaussian function. 
The indistinguishability of the Ne atoms has  been neglected throughout
this work since the Nosanow-Jastrow model is not symmetric under the
exchange of particles. This choice is fairly justified since quantum
effects derived from a correct symmetrization  are not expected to play any
significant role in the solid properties calculated in this work. In fact,
the same conclusion for the same quantities also holds for solid $^4$He, a
solid with a larger quantum behavior.
The parameters of the simulation, namely the number of particles per  
box, time step and target walker population (that is, the mean number of walkers 
along the simulation), have been chosen in order to ensure the correct asymptotic behavior; 
their respective values are: ~$N =256$~, $\Delta\tau = 2.7 \cdot 10^{-4}$~K$^{-1}$ and $n_{w} = 260$~.

At each density, finite size effects have been corrected by including the
tails of the kinetic and potential energies into the total energy,  both
estimated assuming $g(r) = 1$ beyond half the length of the simulation box.
This assumption could be too crude for solids (see Fig.~\ref{fig:gr}) and
therefore we have checked the reliability of this approximation in our
system.  To this end, we have carried out some simulations with 500 atoms and
compared the  energetic and structural results with the ones obtained for a
box of 256 particles at the same density. For instance, at a density 
$\rho=0.045$~\AA$^{-3}$ the energies are $E/N=-238.88(4)$~K and $-238.69(8)$~K and the
Lindemann ratios $\gamma_{\rm Ne} = 0.077(1)$ and $0.079(3)$ for $N=256$ and
$500$ particles, respectively. The differences observed are therefore not
significant within our statistical uncertainty and the size corrections are
reasonably included.

\section{Results}
\label{sec:results}

\subsection{Thermodynamic properties}

\begin{figure}[t]
\centering
       { \includegraphics[width=0.45\linewidth]{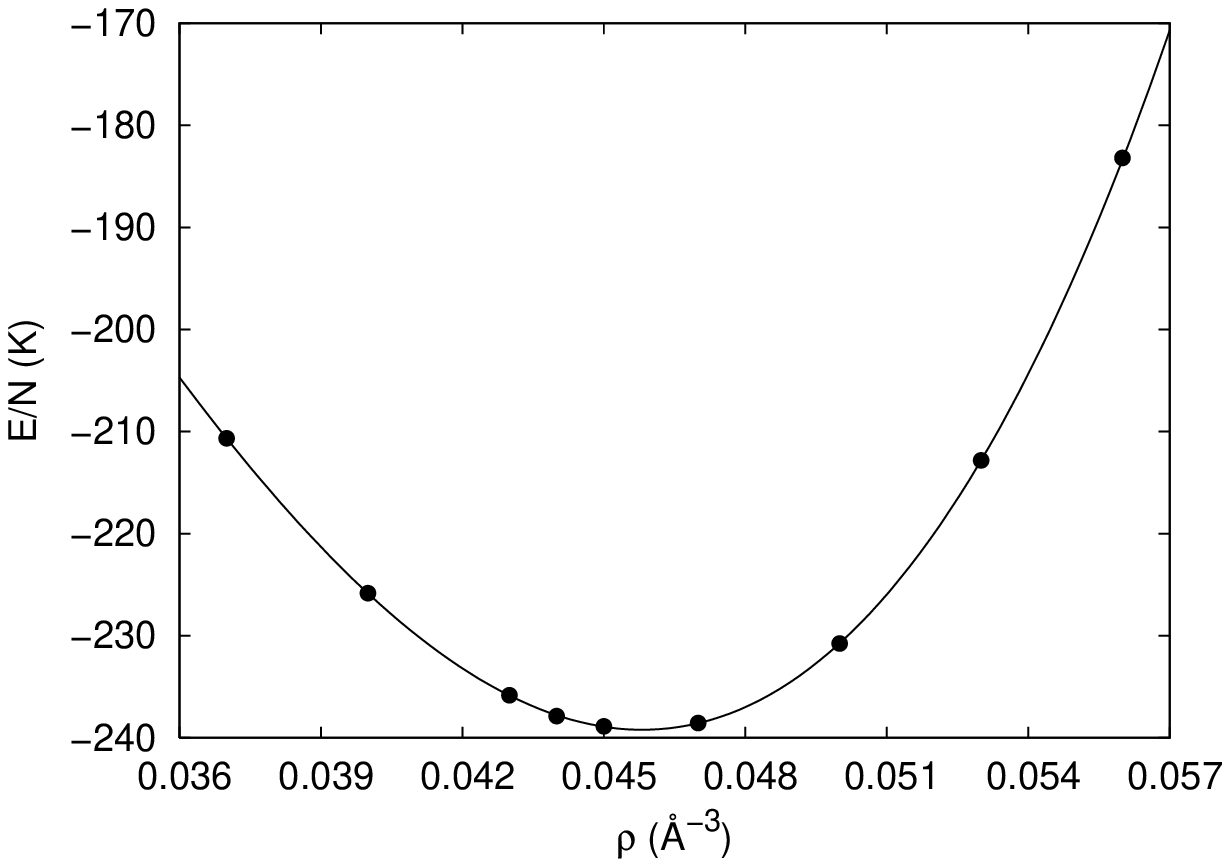} }%
       { \includegraphics[width=0.45\linewidth]{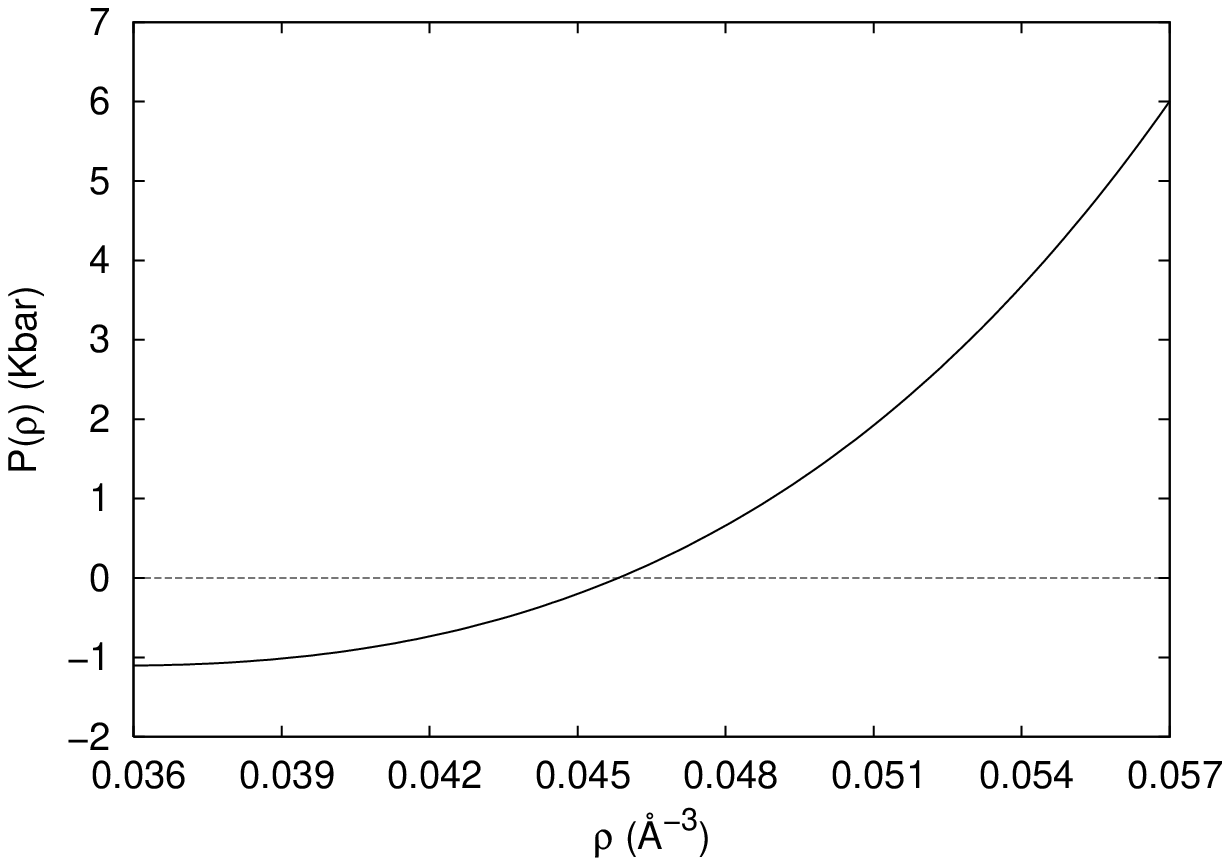} }%
 \vspace{0.0cm}
 \caption{ \emph{Left}: Energy versus density for solid Ne at zero temperature. The solid line corresponds to the polynomial curve of 
	   Eq.~\ref{eq:fitne} and the dots to the calculated DMC total energies per particle, errors bars are smaller than the size
	   of the symbols.    
	   \emph{Right}: Equation of state of solid Ne at zero temperature computed with DMC and the Aziz HFD-B potential.
			 }
\label{fig:eos}
\end{figure}

In Fig.~\ref{fig:eos} (left), we show our results for the total atomic energy of solid Ne at zero temperature. 
The solid line on it corresponds to the polynomial curve, $e\left(\rho\right)=E\left(\rho\right)/N$,
\begin{equation}
\label{eq:fitne} 
e(\rho) = e_{0} + a\left( \frac{\rho - \rho_{0}}{\rho_{0}} \right)^{2} +
	  b\left( \frac{\rho - \rho_{0}}{\rho_{0}} \right)^{3}~,
\end{equation}
which has been fitted to the DMC energies reported in Table~\ref{tab:energies} (solid points in the figure).
The values of the parameters of the best fit are $a = 938(3)$~K, $b = 871(20)$~K, $e_{0} =
-239.21(3)$~K and $\rho_{0} = 0.04582(2)$~\AA$^{-3}$~,   
where $e_{0}$ and $\rho_{0}$ are the equilibrium
energy per particle and density, respectively.
The agreement between our results and experiments is reasonably good for the energy   
$e_{0}^{expt} = -232(1)$~K and the density 
$\rho_{0}^{expt} = 0.044976(3)$~\AA$^{-3}$.~\cite{conville74,batchelder67}

Once $e(\rho)$ is known, it is straightforward to deduce the pressure, $P(\rho)$ (see Fig.~\ref{fig:eos}, right), and compressibility,
$\kappa (\rho)$, of the system at any density through the relations  
\begin{eqnarray}
P(\rho)=\rho^{2} \frac{\partial e(\rho)}{\partial \rho} \nonumber   \\
\kappa (\rho) = \frac{1}{\rho} \frac{\partial \rho}{\partial P}~.
\end{eqnarray}
The compressibility at the equilibrium density obtained so is $\kappa_{0} = 0.084(4)$~Kbar$^{-1}$,
which compares excellently to the experimental value  $\kappa_{0}^{expt}= 0.089(2)$~Kbar$^{-1}$.~\cite{batchelder67}
  
An interesting magnitude in the study of condensed phase systems is the spinodal density,
$\rho_{S}$, which is the thermodynamical limit for the system to remain in homogeneous phase.
At this density, the relation $\partial P\slash \partial \rho = 0$~ is fulfilled,
which is equivalent to require infinite compressibility or zero speed of sound in the system. Our prediction for $\rho_{S}$  
is $0.03575(5)$~\AA$^{-3}$, which corresponds to a pressure $P(\rho_{S}) = -1.102(4)$~Kbar.
In Ref.~\onlinecite{herrero03}, Herrero presents a comprehensive study of solid Ne at negative pressures 
by means of the PIMC method. The author modelizes the interatomic interactions with the L-J potential 
and estimates the pressure at the spinodal density and zero temperature by means of 
a linear fit to the squared bulk modulus with respect to pressure; he obtains $P(\rho_{S})^{PIMC} = -0.91$~Kbar    
and $\rho_{S}^{PIMC} = 0.0356$~\AA$^{-3}$~. 
The disagreement between this and our value for $P(\rho_{S})$ can be explained in terms of the
adopted interatomic potential, since small differences in the total energies may develop large 
within successive derivatives.

\begin{table}
\begin{center}
\begin{tabular}{c c c c }
\hline
\hline
$ \rho $ (\AA$^{-3}$) & $ E/N $ & $ E_{p}/N $ & $ E_{k}/N $ \\ 
\hline
$ 0.040 $ &  $  -225.84 (4) $ & $ -256.26 (8) $  &  $ 30.34  (8) $ \\
$ 0.043 $ &  $  -235.83 (4) $ & $ -272.04 (8) $  &  $ 36.23  (8) $ \\
$ 0.044 $ &  $  -237.88 (4) $ & $ -276.50 (8) $  &  $ 38.57  (8) $ \\
$ 0.045 $ &  $  -238.88 (4) $ & $ -279.43 (8) $  &  $ 40.61  (8) $ \\
$ 0.047 $ &  $  -238.55 (4) $ & $ -283.76 (8) $  &  $ 45.17  (8)  $ \\
$ 0.050 $ &  $  -230.76 (4) $ & $ -282.81 (8) $  &  $ 52.17  (8) $ \\
$ 0.053 $ &  $  -212.83 (4) $ & $ -272.15 (8) $  &  $ 59.31  (8) $ \\
$ 0.056 $ &  $  -183.20 (4) $ & $ -249.61 (8) $  &  $ 66.41  (8) $ \\
\hline
\hline
\end{tabular}
\end{center}
\caption{Total, potential and kinetic energies per particle
of solid Ne at absolute zero as computed with DMC and the pure estimator technique.
Energies are in units of K.}
\label{tab:energies}
\end{table}

\subsection{Structural properties}

\begin{figure}[t]
\centerline{
        \includegraphics[width=0.8\linewidth]{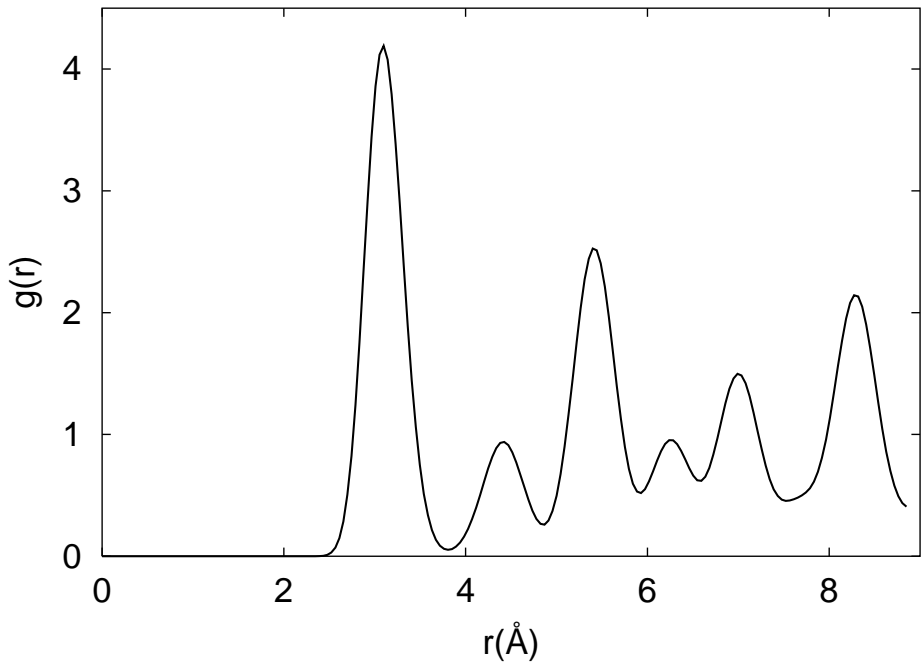}}%
        \caption{Averaged radial pair-distribution function, $g(r)$, of solid Ne at zero temperature and 
		the equilibrium density. }
\label{fig:gr}
\end{figure}

We have explored several structural properties of solid Ne.       
In Fig.~\ref{fig:gr}, we plot the averaged radial pair-distribution function, $g(r)$, which is 
proportional to the probability of finding a particle at a certain distance
$r$ from another. According to what is expected in crystals, $g(r)$ emerges peaked   
with maxima corresponding to the distances between successive shells of atoms within the perfect lattice,
though the peaks broaden with respect to the profiles which are obtained in classical solids.

A characteristic parameter in the study of quantum solids
is the Lindemann's ratio, $\gamma$, which is defined as the ratio between the 
squared root of the mean squared displacement, $\langle {\bf u^{2}}\rangle$, and the 
distance between first nearest neighbours in the perfect crystalline lattice. 
Our estimation of the Lindemann's ratio at the 
equilibrium density (pure estimation) is $\gamma_{\rm Ne} = 0.088(2)$~, which is significantly smaller than in $^{4}$He
($\sim 0.26$) and H$_{2}$ ($\sim 0.18$), but still larger than in classical solids at finite temperature and far from 
melting ($\sim 0.03$).
The corresponding mean squared displacement, $\langle {\bf u^{2}_{\rm Ne}}\rangle$, amounts to $0.077(1)$~\AA$^{2}$~.
In Table~\ref{tab:lindemann}, we quote the value of $\gamma_{\rm Ne}$ at several densities out of the equilibrium. As it is observed therein, 
the general trend of $\gamma_{\rm Ne}$ is to reduce when the density is increased; this behavior is easily understood in terms
of gain of cohesion energy, which must balance with the increasing of kinetic energy of the system arising from atomic 
localization.

\begin{figure}[t]
\centerline{
        \includegraphics[width=0.8\linewidth]{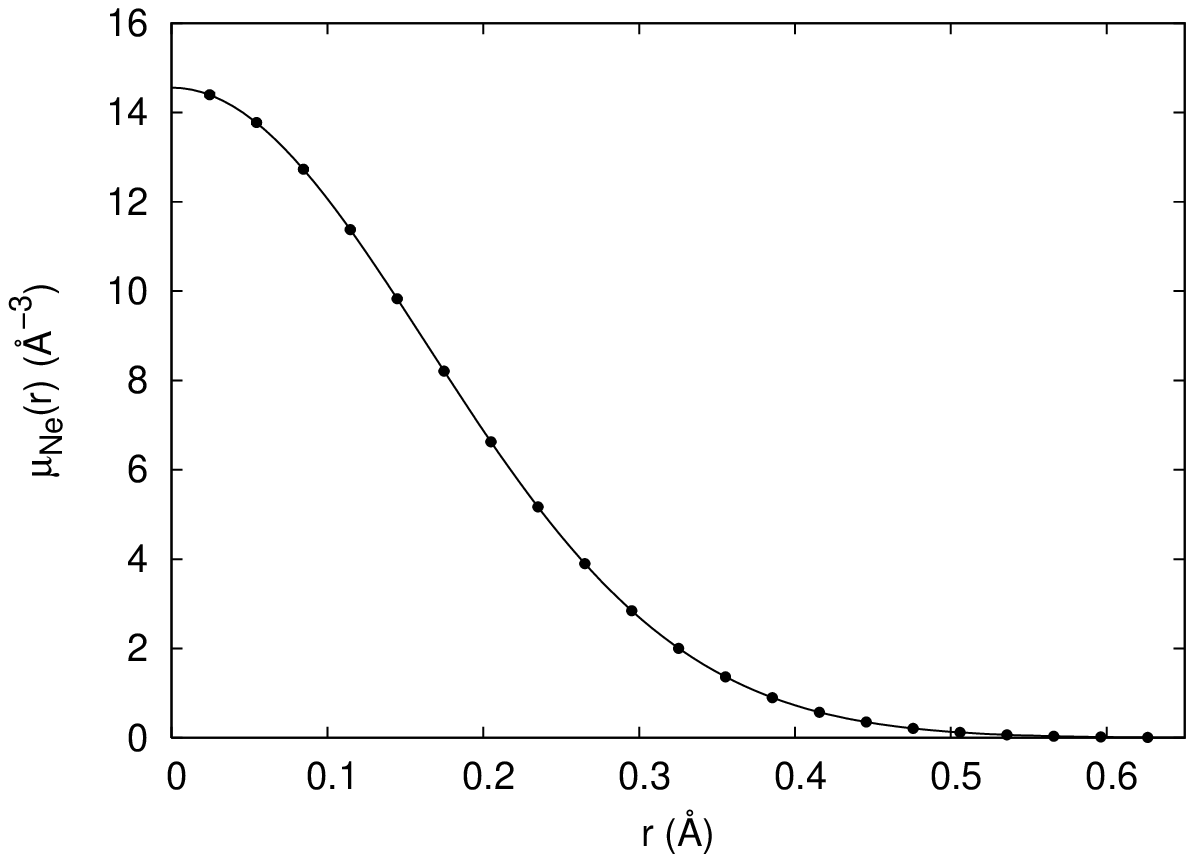}}%
        \caption{ Atomic averaged density profile, $\mu(r)$, of solid Ne at zero temperature and
		 the equilibrium density. }
\label{fig:dprof}
\end{figure}

Aimed to characterize the spatial distribution of the atoms around the equilibrium positions in solid Ne, 
we have calculated the atomic density profile function (averaged for all directions), $\mu(r)$, and kurtosis, $\zeta_{Q}$~.
The averaged atomic density profile function, $\mu(r)$, yields the probability of finding a particle at a distance 
within the interval $\left(r, r+dr \right)$ from any arbitrary site of the lattice.
According to this definition, the mean squared displacement, $\langle {\bf u^{2}}\rangle$, can be obtained as 
\begin{equation}
\label{eq:dprof}
\langle {\bf u^{2}} \rangle = 4\pi\int_{0}^{\infty} \mu(r) r^{4} dr ~. 
\end{equation} 
In Fig.~\ref{fig:dprof}, we plot $\mu(r)$ at the equilibrium density (dots), together with the Gaussian curve that we have adjusted 
to it (solid line). 
To check the reliability of this fit, we have assumed the Gaussian curve in Eq.~(\ref{eq:dprof}), instead of $\mu(r)$,   
and then recalculated $\langle {\bf u^{2}} \rangle$.
Proceeding so, we obtain $0.079(1)$~\AA$^{2}$ which agrees perfectly with the direct calculation $0.077(1)$~\AA$^{2}$~.
Next, we compute $\zeta_{Q}$ in several directions of the cubic cell so as to discern whether
the atoms distribute isotropically in average or not around the sites.  
The kurtosis is defined as
\begin{equation}
\label{eq:kurtosisne}
\zeta_{(ijk)} = \frac{\langle {\bf u^{4}}_{(ijk)}\rangle }{ \langle {\bf u^{2}}_{(ijk)}\rangle^{2} } - 3 ~,
\end{equation}
where ${\bf u}_{(ijk)}$ are the projections of the position vectors which relate each lattice site to its
nearest particle along the $(ijk)$ direction (Cartesian basis).
As it is well-known, if the atomic density distribution over the equilibrium positions is of Gaussian type   
the kurtosis is null.
In the case of solid Ne, we have obtained $\zeta_{(100)} = 0.0078(63)$ and $\zeta_{(010)}= 0.0062(59)$, which indeed might be regarded as values
compatible to zero. Additional results for $\zeta_{Q}$ obtained with the pure estimator technique are quoted in Table~\ref{tab:lindemann}.

\begin{table}
\begin{center}
\begin{tabular}{ c c c c }
\hline
\hline
$ \rho $ (\AA$^{-3}$) & $ \gamma_{\rm Ne} $ & $ \zeta_{(100)} $ & $ \zeta_{(010)}$ \\ 
\hline
$ 0.040 $ &  $ 0.099(2)  $ & $ 0.017(12) $  &  $ 0.012(14) $ \\
$ 0.043 $ &  $ 0.092(2)  $ & $ 0.000(8)  $  &  $ -0.001(8) $ \\
$ 0.044 $ &  $ 0.091(2)  $ & $ 0.000(10) $  &  $ 0.000(10) $ \\
$ 0.045 $ &  $ 0.087(2)  $ & $ -0.006(7) $  &  $ -0.014(7) $ \\
$ 0.047 $ &  $ 0.086(2)  $ & $ 0.000(20) $  &  $  0.000(10)$ \\
$ 0.050 $ &  $ 0.083(2)  $ & $ 0.000(10) $  &  $ -0.010(10)$ \\
\hline
\hline
\end{tabular}
\end{center}
\caption{ Lindemman's ratio, $\gamma_{\rm Ne}$, and kurtosis, $\zeta_{Q}$ of 
solid Ne at different densities close to equilibrium.}
\label{tab:lindemann}
\end{table}
  
\subsection{Kinetic energy and momentum distribution}

\begin{figure}[t]
\centerline{
        \includegraphics[width=0.8\linewidth]{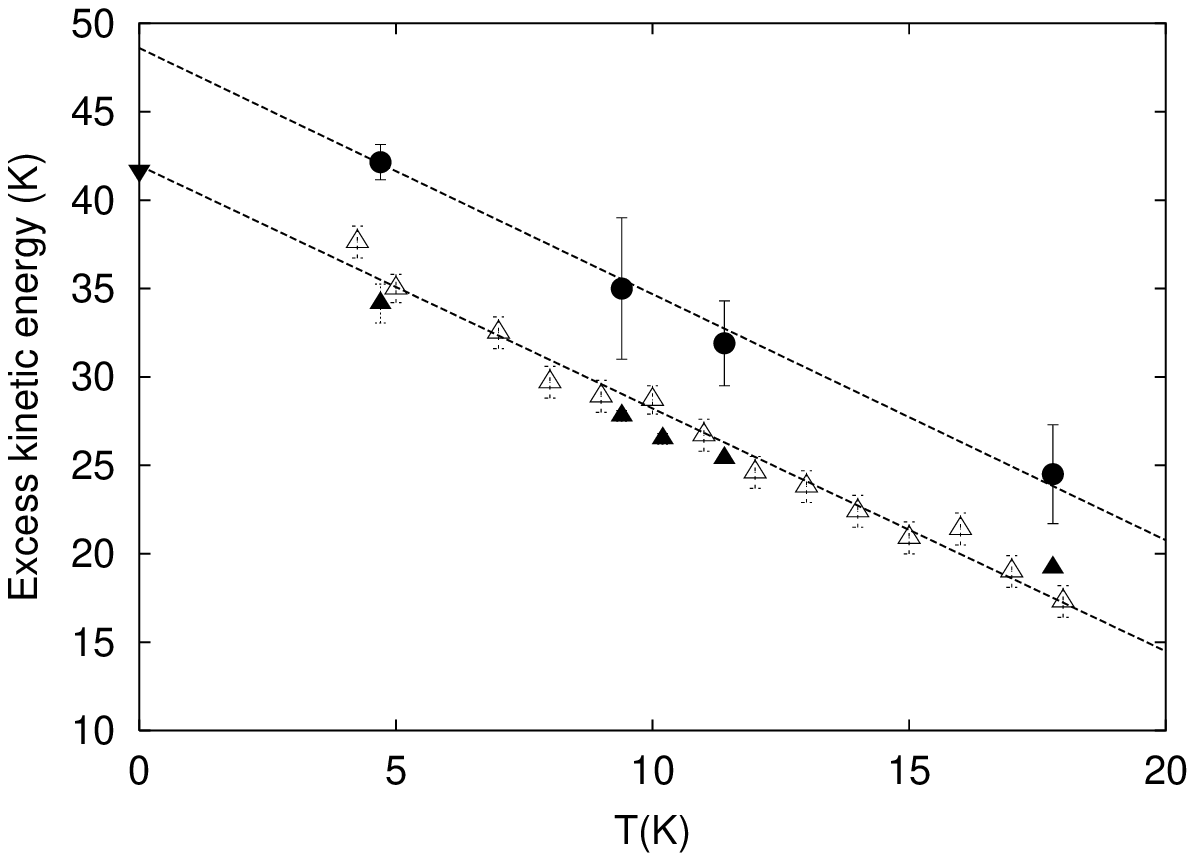}}%
        \caption{Excess atomic kinetic energy of solid Ne at low temperatures. 
		 Experimental data of Ref.~\onlinecite{timms03} are represented by $\vartriangle$~,
		 measurements of Ref.~\onlinecite{peek92} by $\bullet$~, PIMC estimations of Ref.~\onlinecite{timms96}
		 by $\blacktriangle$ and our ground state estimation by $\blacktriangledown$ (in the ordinate axis).
		 The lines in the plot correspond to linear fits to the experimental data of Refs.~\onlinecite{timms03,peek92}~.}
\label{fig:ekin}
\end{figure}

\begin{figure}[t]
\centerline{
        \includegraphics[width=0.8\linewidth]{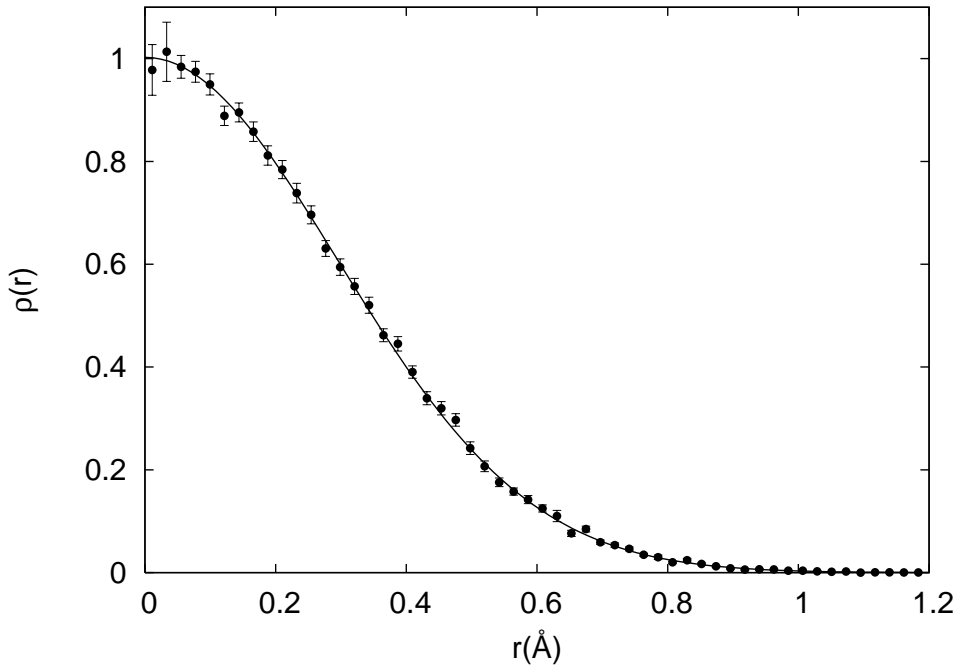}}%
        \caption{One-body density matrix of solid Ne at the equilibrium density.
		 The solid line in the figure corresponds to the Gaussian curve that we have fitted to the results.}
\label{fig:onebody}
\end{figure}

\begin{figure}[t]
\centerline{
        \includegraphics[width=0.8\linewidth]{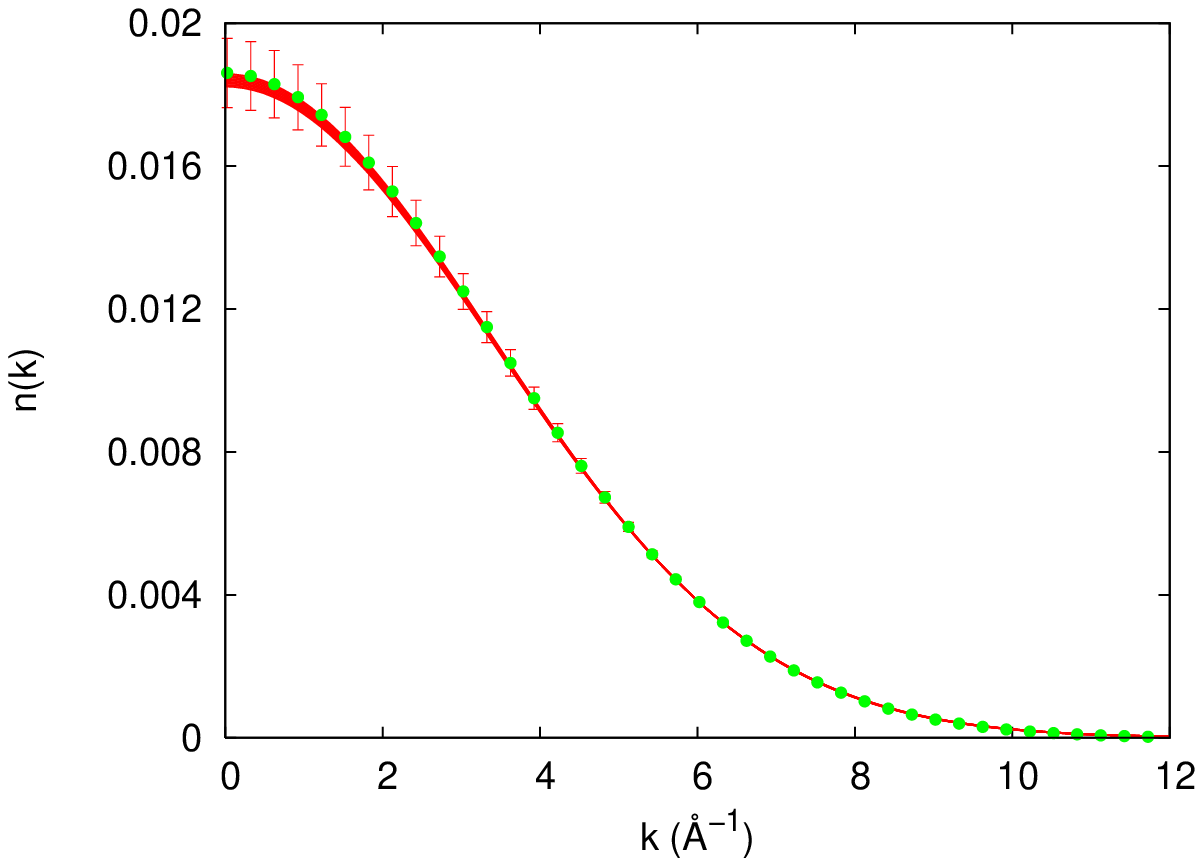}}%
        \caption{(Color online) Momentum distribution of solid Ne at the equilibrium density (green dots and bars). 
		 The solid line in the figure corresponds to the Fourier transform of the Gaussian curve previously fitted 
	         to $\varrho (r)$ (the width of the line represents the uncertainty of the fit).}
\label{fig:momentum}
\end{figure}

In Table~\ref{tab:energies}, we summarize the value of the atomic ground-state kinetic and potential energies of solid Ne
near equilibrium ($P \sim 0$). All the $E_{p}$ and $E_{k}$ results have been computed
within the pure estimator technique and DMC, thus any possible errors associated to them are of statistical kind or stem from
the modelization of the interatomic interactions.
In particular, we have estimated $E_{k} = 41.51(6)$~K at the equilibrium density.
In Fig.~\ref{fig:ekin}, we plot the values of the excess kinetic energy of solid Ne, defined as
$E_{exc} = E_{k} - \left(3/2\right) T$, as measured by Peek~\cite{peek92} and Timms {\it et al.}~\cite{timms03} 
within the temperature range $4 - 20$~K. Therein, we also include estimations of $E_{exc}$ as obtained with PIMC  
over the same $T$-interval, together with our ground-state result which is located at the ordinate axis.
By performing linear fits to the excess kinetic energy, it is shown that our ground-state prediction
is in very good agreement with Timms's measurements~\cite{timms03} and the PIMC estimations,~\cite{timms96} 
whereas not so with Peek's results.~\cite{peek92} 
The causes for this disagreement may be explained, as it has been suggested elsewhere,~\cite{neumann02,timms03}  
in terms of systematic experimental errors, since the temperature dependence of $E_{exc}$ obtained by Peek and 
co-workers appears to coincide with Timms' results. 
A likely explanation can rely on the range of neutron momentum transfers involved
in those first DINS experiments, about two orders of magnitude less intense than in posterior measurements, which might not be sufficiently
large so as to reach the high $Q$-regime required for the impulse approximation of the dynamic structure factor to be
valid.~\cite{timms03}

Another physically rich quantity in the study of quantum liquids and solids is the one-body density matrix, $\varrho ({\bf r},{\bf r'})$,
which is defined as
\begin{equation}
\varrho({\bf r},{\bf r'}) = \langle \Phi_{0} | \widehat{\psi}^{\dagger}({\bf r}) \widehat{\psi}({\bf r'}) | \Phi_{0} \rangle ~,
\label{eq:onebody}
\end{equation}
where $\widehat{\psi}({\bf r'})$ and $\widehat{\psi}^{\dagger}({\bf r})$ are, respectively, the field operators which destroy a particle from 
position ${\bf r'}$ and create one at position ${\bf r}$ and $\Phi_{0}$ is the ground-state wave function. 
In boson systems the asymptote $\lim_{r \to \infty} \varrho(r)$ provides the condensate fraction of the associated homogeneous system
$n_{0}$. The Fourier transform of $\varrho (r)$ is directly the atomic momentum distribution,
\begin{equation}
n({\bf k}) = \rho \int d{\bf r}~ e^{i {\bf k}\cdot {\bf r}}~ \varrho (r) ~.
\label{eq:momentumdist}
\end{equation}
In the Quantum Monte Carlo formalism, the one-body density function can be estimated 
by averaging the coordinate operator
$A({\bf r},{\bf r_{1}},...,{\bf r_{N}}) \equiv  \psi_{T}({\bf r_{1}}+{\bf r},{\bf r_{2}},...,{\bf r_{N}}) / \psi_{T}({\bf r_{1}},
{\bf r_{2}},...,{\bf r_{N}})$ within customary DMC (known as mixed estimation, $\varrho_{mix}(r) = \langle A(r)\rangle_{DMC}$).~\cite{moroni97}
However, a more accurate evaluation of $\varrho (r)$, known as extrapolated estimation, is given by the expression
\begin{equation}
\varrho(r) = 2 \varrho_{mix}(r) - \varrho_{VMC}(r)~,
\label{eq:extrapolone}
\end{equation}  
where $\varrho_{VMC}(r)$ results from averaging $A({\bf r},{\bf r_{1}},...,{\bf r_{N}})$ within variational Monte Carlo.
In Fig.~\ref{fig:onebody}, we plot our results for $\varrho (r)$ as given by Eq.~(\ref{eq:extrapolone}).
In the same figure, we also enclose the Gaussian curve, $G(r) = e^{-br^{2}}$ (given that $\varrho(0) = 1$), which best fits to our
calculations, with an optimal parameter value $b = 5.743(36)$~\AA$^{-2}$. 
In order to test the quality of this fit (which in the reduced chi-squared test gives the value $0.99$), we have
calculated the atomic kinetic energy of solid Ne through the formula
\begin{equation}
E_{k} = -\left[ \frac{\hbar^{2}}{2m_{\rm Ne}}\nabla^{2} \varrho (r)\right]_{r = 0} ~,
\label{eq:gaussianproof}
\end{equation}
but assuming $G (r)$ instead of $\varrho (r)$. In fact, it may be shown that Eq.~(\ref{eq:gaussianproof})  
derives from the kinetic-energy sum rule
\begin{equation}
E_{k} = \frac{\hbar^{2}}{2m_{\rm Ne}} \frac{1}{\left(2\pi\right)^{3}\rho}\int d{\bf k}~k^{2}~n(k)~.
\label{eq:sumrule}
\end{equation}
Proceeding so, we have obtained $E_{k} = 41.43(26)$~K, which fully agrees with the direct estimation $41.51(6)$~K. This finding 
allows us to conclude that $\varrho (r)$ in solid Ne at $T = 0$ can be well considered Gaussian-shaped at all effects.  

We have also computed the atomic momentum distribution of solid Ne by taking the Fourier transform of $\varrho (r)$
over a set of $k$-vector points, as expressed in Eq.~(\ref{eq:momentumdist}). In Fig.~\ref{fig:momentum}, we plot the results
of these calculations (dots) and additionally the Fourier transform of the aforegiven Gaussian fit to $\varrho (r)$ (solid line
with width signalizing the associated uncertainty). Obvioulsy, once $\varrho (r)$ has proved Gaussian, $n(k)$ turns out to be of the 
same kind.

\subsection{Degree of anharmonicity}

The Self Consistent Phonon approach~\cite{glyde94} (SCP) has proved very accurate
in characterizing solids in the middle way between classical and quantum behavior. 
Very essentially, this theory makes the assumption of particles coupled harmonically with frequencies  
and modes depending on the crystal symmetry and lattice parameter and which are determined through a self-consistent
procedure. A simplified version of this method is the Self Consistent Average Phonon approach
(SCAP),~\cite{shukla81,paskin82} which adopts the expressions of SCP  
but replacing the summation over the different vibrational
frequencies by an averaged one, namely the Einstein frequency, $\Omega_{0}$.
Despite this crude simplification, the agreement between measurements and  
SCAP results for rare gase solids like Ar, Kr and also Ne is excellent in what concerns
thermodynamic properties (isothermal bulk modulus, specific heat, etc.).~\cite{mohazzabi87}
However, we want to know to what extent harmonic assumptions in solid Ne are accurate enough for deriving   
microscopic properties of its ground state.
With this aim, we have calculated the atomic kinetic energy and mean squared displacement within SCAP through the
formulas 
\begin{equation}
\Omega_{0}^{2} =
\frac{1}{3 m_{\rm Ne} N } \left \langle \sum_{i=1}^{N}  
\mbox{\bf $\nabla$}_i^2
V_{2}({\bf r}) \right \rangle \ 
\label{eq:omega2}
\end{equation}
and
\begin{eqnarray} 
\label{eq:scapmeankin}
\langle {\bf u}^{2}
\rangle^{\rm (SCAP)} & =& \frac{3 \hbar}{2 m_{\rm Ne} \Omega_{0}}
~, \nonumber \\ \langle E_{k} \rangle^{\rm (SCAP)} & = & \frac{1}{2} ~
m_{\rm Ne} \langle {\bf u}^{2} \rangle^{\rm (SCAP)}  \Omega_{0}^2 = \frac{3}{4} \hbar \Omega_{0}~, 
\end{eqnarray} 
where $V_{2}(r)$ is the Aziz HFD-B pair potential.
We first compute the exact value of $\Omega_{0}$ 
with the pure estimator technique within the DMC approach, and then calculate  
the value of expressions (\ref{eq:scapmeankin}).
The results that we have obtained are, $\hbar\Omega_{0} = 62.04(1)$~K, $\langle{\bf u}^{2}_{\rm Ne}\rangle^{SCAP} =
0.058(3)$~\AA$^{2}$ and $\langle E_{k}\rangle^{SCAP} = 46.5(1)$~K, which disagree noticeably from the DMC values
$\langle{\bf u}^{2}_{\rm Ne}\rangle = 0.077(1)$~\AA$^{2}$ and $\langle E_{k}\rangle = 41.51(6)$~K. This outcome 
reveals that crude simplifications made on the vibrational properties of solid Ne may lead to important inaccurracies 
on the quantum description of such crystal. 

In a further step, we have devised an harmonic model~\cite{wallace72} in which
the interaction between particles is pairwise and reads
\begin{equation}
V_{2}^{\rm harm}(r_{ij}) = V_{2}(r_{0,ij}) +
\frac{1}{2} \left( {\bf u}_i -
{\bf u}_j \right)^{\rm T} \left( \frac{ \partial ^2 V_{2} }{ \partial
{\bf r}_{ij} \partial {\bf r}_{ij} } \right)_{ r_{ij}= r_{0,ij}}\left(
{\bf u}_i - {\bf u}_j \right)~,
\label{eq:harmonichal}
\end{equation}
where $V_{2} (r)$ is the Aziz HFD-B interaction, ${\bf u}_{i}$ is defined as ${\bf r}_{i} - {\bf R}_{i}$, and the terms
$ V_{2}(r_{0,ij})$ and $\left( \frac{ \partial ^2 V_{2} }{ \partial {\bf r}_{ij} \partial {\bf r}_{ij} } \right)_{ r_{ij}= r_{0,ij}}$ 
in the right side of Eq.~(\ref{eq:harmonichal}) are evaluated, only once, for the atoms in the perfect crystal configuration
($r_{0,ij} \equiv |{\bf R}_{i} - {\bf R}_{j}|$) .
This approach is equivalent to assume the pair of atoms $i$ and $j$ coupled through an harmonic spring of constant equal to 
the second derivative of $V_{2}(r)$ evaluated at the equilibrium distance $r_{0,ij}$~. 
Within DMC and with the pure estimator technique, we have computed the exact ground-state total and kinetic energies of this model, arriving
at the values, $e_{0}^{\rm harm} = -251.35(4)$~K and $E_{k}^{\rm harm} = 35.1(3)$~K, which differ notably from the
results obtained with the full Aziz HFD-B interaction.  

The relative failure of the previous approximations allow us to conclude that traditional harmonic approximations in solid Ne are not adequate for 
an accurate evaluation of its microscopic properties. Aimed to yield a rough estimation of the degree
of anharmonicity of solid Ne, and to finalize with this section, we now compare solid Ne with solid $^4$He, the most anharmonic 
among all the crystals, by invoking the Debye model.
In the Debye approach for solids, particles are assumed as non-interacting quantum harmonic oscillators which vibrate with frequencies within
a spectrum that is top-bounded by the Debye frequency, $\omega_{D}$. Consequently, the atomic kinetic energy is expressed as    
$E_{k}^{D} = \left(9/16\right) \Theta_{D}$, where $\Theta_{D}$ is the Debye temperature and is equal to $\hbar \omega_{D}$.     
It is readily shown that $\Theta_{D} = 9 \hbar^{2} / 4 m \langle{\bf u}^{2}\rangle$~, which in the case of solid Ne at equilibrium
turns out to be $70.3(9)$~K (here, we have used the value $\langle{\bf u}^{2}_{\rm Ne}\rangle = 0.077(1)$~\AA$^{2}$), which in turn leads to 
$E_{k}^{D} = 39.5(5)$~K. Next, we define the dimensionless parameter, $\Gamma \equiv 1.0 - \left(E_{k}^{D}/E_{k}\right)$, which in fact   
vanishes for the case of a pure harmonic solid (Debye model) and it progressively increases towards unity as anharmonic effects develop
larger. For solid Ne and $^{4}$He at their respective zero-temperature equilibrium volumes, we assess the values $\Gamma_{\rm Ne} = 0.05$
and $\Gamma_{\rm He} = 0.44$, where for helium we have used the data found in Ref.~\onlinecite{cazorla07}. By comparing these two figures,
one could claim that anharmonic effects in solid Ne are about one order of magnitude less substantial than in $^{4}$He.

\section{Discussion and Conclusions}
\label{sec:discussion}
In this work, we report the calculation of the ground-state atomic kinetic energy, one-body density matrix and momentum 
distribution of solid Ne by means of the DMC method and the realistic Aziz HFD-B pair-potential. Our approach is proved to perform
notably for this crystal, as it is shown by the very good overall agreement obtained with respect to thermodynamic experimental data. 
Our value for the atomic kinetic energy of solid Ne at the equilibrium volume, $E_{k} = 41.51(6)$~K, is in accordance with the low-temperature
experimental data found in Refs.~\onlinecite{timms96,timms03} and also with previous PIMC calculations performed with the L-J and Aziz
HFD-C2 pairwise interactions.~\cite{cuccoli93,timms96,neumann02}
However, our result does not agree with the results obtained by Peeks and co-workers (previous to Timms' work) based also on deep-inelastic neutron
scattering measurements.
We have calculated the one-body density function of solid Ne and shown that it perfectly fits to a Gaussian curve. 
Consequently, the atomic momentum distribution, which is evaluated by performing the Fourier transform of $\varrho (r)$, is of the same kind.
Interestingly, Withers and Glyde~\cite{withers07} have shown very recently by means of simple models that the deviation of $n(k)$ 
from a Gaussian pattern in quantum solids may arise by effect of anharmonicity and/or the introduction of atomic exchanges.
We have checked that anharmonic effects in the ground-state of solid Ne are relevant by calculating some of its microscopic properties
within traditional harmonic schemes and quoting significant discrepancies with respect to the full quantum results. 
It is noted that we have not attempted to include atomic exchange effects in the present work since \emph{a priori}  
and very reasonably, these are not expected to play any substantial role in the ground-state of 
solid Ne (contrarily to what may occur in $^{4}$He, for instance). 
Even so, we do not appreciate, within the statistical uncertainty, any deviation from a Gaussian pattern in 
the $\varrho (r)$, or equivalently $n(k)$, of solid Ne, therefore, the degree of anharmonicity of Ne at zero
temperature may be regarded as fairly moderate.

\acknowledgments
We acknowledge financial support from DGI (Spain) Grant No. FIS2005-04181 and Generalitat de Catalunya Grant No. 2005GR-00779.

\end{document}